\documentclass[utf8]{article} %
\usepackage[onehalfspacing]{setspace}

\usepackage[T1]{fontenc}
\usepackage[utf8]{inputenc}
\usepackage{amsmath,scalerel}
\usepackage{amsfonts}
\usepackage{gauss}
\usepackage{authblk}
\usepackage{biblatex}
\usepackage{float}
\usepackage{svg}
\usepackage{chemformula}
\usepackage{hyperref}
\usepackage{nth}
\usepackage{siunitx}
\usepackage{geometry}
\usepackage{todonotes}
\usepackage{authblk}

\usepackage{xcolor} 
\definecolor{LightGray}{gray}{0.5}
\definecolor{bg}{rgb}{0.95,0.95,0.95}
\usepackage{physics}
\hypersetup{
    colorlinks=true,
    linkcolor=black,
    filecolor=magenta,  %
    urlcolor=blue,%
    citecolor=black,
}

\renewcommand{\d}{^\dagger}
\newcommand{\vd}{^{\vphantom\dagger}}

\newcommand{\infint}{\int _{-\infty} ^\infty}
\renewcommand{\eqref}[1]{equation (\ref{#1})}
\newcommand{\wt}{\widetilde}

\newcommand{\up}{\uparrow}
\newcommand{\ra}{\rightarrow}

\newcommand{\down}{\downarrow}

\newcommand{\eps}{\varepsilon}

\newcommand{\E}[1]{\expval{#1}}
\newcommand{\p}{\partial}

\newcommand\correspondingauthor{\thanks{Corresponding author: mabr@dtu.dk}}
\def\helveticabold{\fontfamily{phv}\bfseries\selectfont}
\newcommand{\keyFont}{\fontsize{8}{11}\helveticabold }

\def\correspondance#1{\global\def\@correspondance{#1}}

\title {Manipulation of magnetization and spin transport in hydrogenated graphene with THz pulses} 
\author[1]{Jakob Kjærulff Svaneborg}
\author[1]{Aleksander Bach Lorentzen} 
\author[1,2]{Fei Gao}
\author[1,3]{Antti-Pekka Jauho}
\author[1]{Mads Brandbyge\correspondingauthor}
\affil[1]{\small Department of Physics, Technical University of Denmark, DK-2800 Kongens Lyngby, Denmark}
\affil[2]{Donostia International Physics Center (DIPC), 20018 Donostia-San Sebasti\'an, Spain}
\affil[3]{Center for Nanostructured Graphene (CNG), Technical University of Denmark, DK-2800 Kongens Lyngby, Denmark}

\begin{document}
\onecolumn

\date{August 7, 2023}

\maketitle

\begin{abstract}
Terahertz (THz) field pulses can now be applied in Scanning Tunnelling Microscopy (THz-STM) junction experiments to study time resolved dynamics. The relatively slow pulse compared to the typical electronic time-scale calls for approximations based on a time-scale separation. Here, we contrast three methods based on non-equilibrium Green's functions (NEGF): (i) the steady-state, adiabatic results, (ii) the lowest order dynamic expansion in the time-variation (DE), and (iii) the auxiliary mode (AM) propagation method without approximations in the time-variation. We consider a concrete THz-STM junction setup involving a hydrogen adsorbate on graphene where the localized spin polarization can be manipulated on/off by a local field from the tip electrode and/or a back-gate affecting the in-plane transport. We use steady-state NEGF combined with Density Functional Theory (DFT-NEGF) to obtain a Hubbard model for the study of the junction dynamics. Solving the Hubbard model in a mean-field approximation, we find that the near-adiabatic first order dynamical expansion provides a good description for STM voltage pulses up to the 1 V range.
\end{abstract}

{\keyFont{ \subsection*{Keywords}  THz spin-electronics,  time-dependent transport, graphene magnetism, non-equilibrium Green's functions (NEGF), Wigner representation, DFT-NEGF}}

\pagebreak
\section{Introduction}
Recently it has become possible to extend Scanning Tunneling Microscopy (STM) studies to the time domain by applying strong sub-cycle near-field electromagnetic pulses in the THz regime inside the STM -- so-called THz-STM. This enables studies of time-resolved dynamics of the transport in junctions with sub-{\AA} and sub-ps resolution \cite{Ammerman2021,Cocker2021,Wang2022}. Due to the strong field-confinement at the STM tip electrode the effective voltages between tip and sample can be on the order of 1 Volt \cite{Peller2021}.

The STM-THz experiments pose interesting questions regarding the theoretical description of the dynamics of the electronic subsystem on the time-scale of typical atomic vibrations. The case of atomic-scale junctions in STM under strong bias and coupling to the tip requires the consideration of coupling to semi-infinite electrodes. To this end the non-equilibrium Green's function (NEGF) methods (e.g. \cite{antti-book,stefanucci2013nonequilibrium,bonitz2016quantum}) have been a popular choice, where the electrodes are included via self-energies. However, it is relevant to consider simplifications to the full time-dynamics since the electron dynamics most often nearly adiabatically follow fields in the THz-range. This can be accomplished by the Wigner representation involving a time-scale separation into a slow THz time-scale ($T$), and fast electronic time-scale ($\tau$), typically around a femto-second. This has been considered recently by Honeychurch and Kosov \cite{Honeychurch2019TimescaleFields} where the NEGF Kadanoff-Baym equations (KBEs) are expressed in the so-called Wigner form with an expansion in terms of the derivative of the slow time, $T$, from hereon called "dynamic expansion" (DE).

From a numerical point of view one important advantage of the DE approach is that each discrete time-step will be independent from each other: Quantities like electron density and current are expressed by quantities independent of the other time-steps allowing for computational parallelization over time as opposed to the time-propagation of the KBEs. This is important for efficient calculations based e.g. on first principles methods such as time-dependent density functional theory based on NEGF (TD-DFT-NEGF). 

In this paper we apply the DE method to a realistic STM junction which has received considerable attention, namely STM on a hydrogen adsorbate on graphene. This system is magnetic, but the magnetism can be tuned by the applied tip or gate field resulting in a change in the in-plane spin-transport. Here, we will consider the junction subject to a THz-pulse and use it to benchmark the DE against the full time-dependent NEGF calculation with the auxiliary mode (AM) method, and the steady-state field approximation (zero-order DE). We first use steady-state DFT-NEGF to examine the bias- and gate-dependent electronic structure and in-plane transport in the system. We extract parameters from here to obtain an effective mean-field, Hubbard-model for the junction, which we use for the benchmarking and which serves as a toy-model of a full TD-DFT-NEGF calculation, highlighting the numerical considerations in a self-consistent treatment of the time-dependence within the DE approach. Our results show how the computationally efficient DE method is able to capture the main features in the THz dynamics of the system, in particular that the THz-bias pulse gives rise to fast switching dynamics of the magnetism, generating higher-harmonic dynamics in the occupations and currents which, on the other hand, are not captured well by the steady-state adiabatic approximation.

\section{System and DFT calculations}

\begin{figure*}
\includegraphics[width=\textwidth]{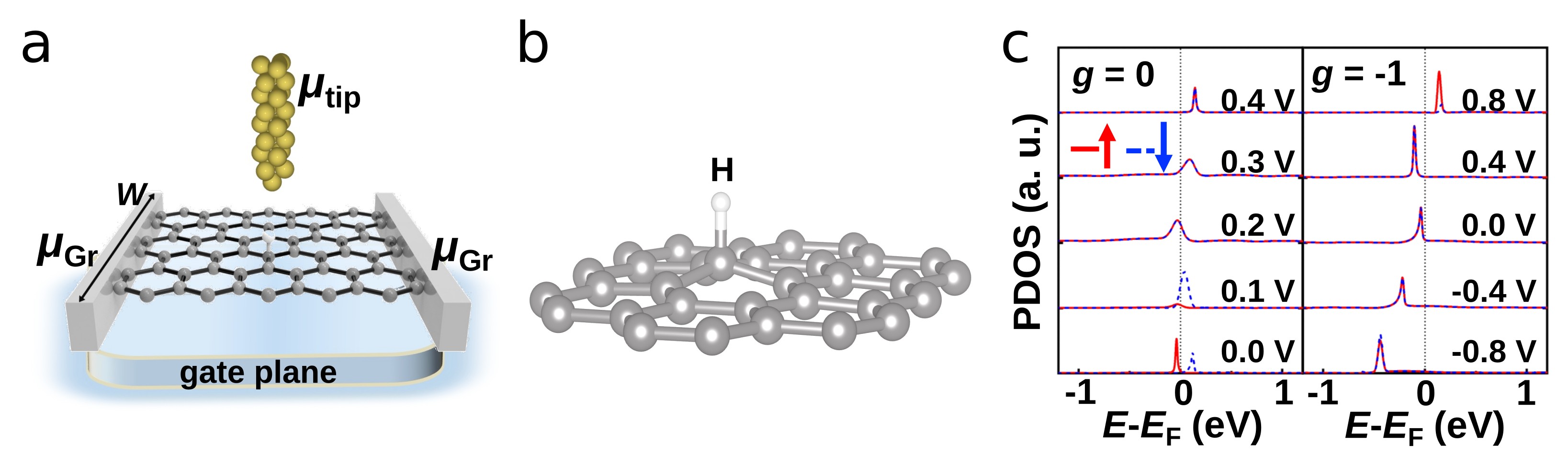}
\caption{\label{fig:H} (a) Schematic of the three-terminal hydrogen on graphene (H/Gr) device setup with a charge back-gate plane. The $\mu_\text{tip}$ and $\mu_\text{Gr}$ are chemical potentials of the gold tip and graphene electrodes. Here, $W$ is the width of the device. (b) Local atomic structure of hydrogen atom on graphene showing an out-of-plane buckling of the C atom below the H (${\rm sp}^3$ hybridization). 
(c) Density of states projected on the H atom, the midgap states, with $\emph{g}=0$ and $\emph{g}=-1$ gating in units of $10^{13}{\rm cm}^{-2}$, respectively. Solid red and dashed blue lines represent spin up and down populations, respectively. $E_F$ refers to the Fermi energy of the system in equilibrium i.e. without bias.} 
\end{figure*}

Magnetism of graphene can be created by point defects like adsorption of isolated hydrogen (H/Gr) or a vacancy, owing to the creation of unpaired $\pi$ electrons around the defects  \cite{H}. The electronic density of graphene is highly tunable by electrostatic gating  \cite{Novoselov2004} which makes it interesting to consider how this magnetism may be tunable either by a global back-gate or a local gating such as the voltage on the STM tip, see Fig.~\ref{fig:H}a-b.

First, we consider steady-state DFT-NEGF calculations where we employ a three-terminal device setup with a gate plane, encompassing left and right graphene electrodes (with width $W$), and an additional gold tip electrode above the H atom, schematically shown in Fig.~\ref{fig:H}a. The calculations were performed using the SIESTA/TranSIESTA package  \cite{brandbyge2002density, siesta, transiesta, sisl}  with the GGA-PBE  \cite{PBE} functional for exchange-correlation, a DZP atomic orbital basis-set and an electronic temperature of 50 K (for further details see Gao {\it et al.} \cite{Gao2021ControlGating}).
To model the gate-induced doping of graphene, a gate plane was placed 15~{\AA} underneath the graphene. The gate carries a charge density of $n=\emph{g}\times 10^{13} \,e/\rm{cm}^2$, where $\emph{g}$ defines the gating level, with $\emph{g}<0$ ($\emph{g}>0$) corresponding to $\emph{n}$($\emph{p}$) doping  \cite{gating}.

The local atomic structure of H/Gr in Fig. \ref{fig:H}b displays an out-of-plane buckling of the C atom, leading to a transition from sp$^2$ to sp$^3$ hybridization.
In Fig.~\ref{fig:H}c we show the calculated density of states around the Fermi energy ($E_F$) projected on the H atom (PDOS) without gating, $\emph{g}=0$, and with gate-induced $n$-doping, $\emph{g}=-1$.
The H impurity resonances in the center of graphene’s pseudogap, the midgap peaks, imply a strong interaction between the s orbital of the H atom and the $p_z$ orbital of graphene. Here, the H adatom forms a $\sigma$ bond with the carbon atom next to it, and the $\pi$ bonds are broken. The bonding states are located at $-8$ eV, far below the Fermi level.

We employ a tip-H distance of 4.5~\AA, where there is only a weak connection between tip and H to keep the spin moment at 1 $\mu_\text{B}$ as the initial state at zero bias ($0$ V) and gate. Without gating, the magnetic moment can be switched on/off with the tip voltage induced doping below the tip, as seen by the disappearance of the spin-splitting in the PDOS. On the other hand, we can also turn the magnetism "off" for a $n$-doped graphene by applying gating ($\emph{g}$ = -1), where we see a non-spin-split, fully occupied peak for zero tip bias ($0$ V). In this situation the magnetism/spin-splitting reappears when we apply a $0.8$ V tip bias counteracting the gate doping. Thus, it is possible to manipulate the spin by either the STM tip bias, or by a global field from a gate plane  \cite{Gao2021ControlGating}. The main driver for this behavior is the electronic occupation of the carbon below the hydrogen.

\begin{figure*}
\includegraphics[width=\textwidth]{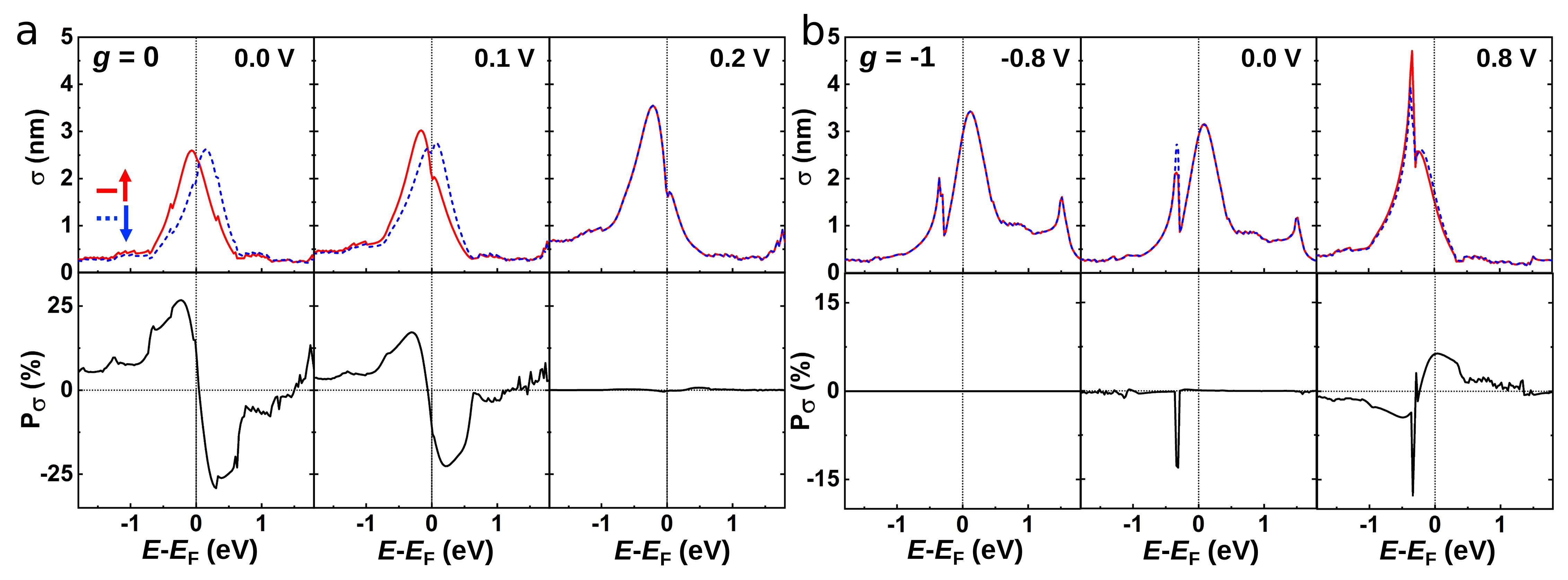}
\caption{\label{fig:CS} In-plane spin-dependent elastic scattering cross section ($\sigma$) as a function of energy. In the bottom panel, the corresponding spin polarization ($P_\sigma$) without (a) $\emph{g}=0$ and (b) with $\emph{g}=-1$ gate-induced $n$-doping. Solid red and dashed blue lines represent spin up and spin down states, respectively.} 
\label{fig.crosssection}
\end{figure*}

Next we address the effects of the field and spin-manipulation on the in-plane transport in H/Gr. To this end we consider the energy-dependent scattering cross section, $\sigma(E)$  \cite{CS},  corresponding to the "shadowing" length (area) caused by the point defect in 2D (3D) conductors. We can estimate $\sigma(E)$ from the electron transmission functions for the pristine and defected systems, $\emph{T}_0$ and $\emph{T}$, respectively  \cite{SCS-1,SCS-2}, 
\begin{equation}
   \sigma(E) = W\frac{T_0(E)-T(E)}{T(E)},
   \label{eq.crosssection}
\end{equation}

where $\emph{W}$ is the width of the device as shown in Fig.~\ref{fig:H}a. In Fig.~\ref{fig:CS}a we see two peaks around $E_F$ caused by the resonant, spin-split H resonances, cf. the PDOS in Fig.~\ref{fig:H}c, yielding a significant in-plane transport spin polarization, $P_\sigma = (\sigma_\uparrow - \sigma_\downarrow)/(\sigma_\uparrow + \sigma_\downarrow)$. The cross section we obtain here is close to the one obtained for H on graphene nanoribbons  \cite{CS}. 
In accordance with the PDOS in Fig. ~\ref{fig:H}c we observe in Fig.~\ref{fig:CS}a how the spin splitting and transport spin polarization vanish when applying a tip voltage.
On the other hand, in Fig.~\ref{fig:CS}b we show how the in-plane transport spin polarization can be turned "on" using the counteracting tip bias for the gate-induced $n$-doped system. In this case we also observe a dip in the cross section around -0.3 eV below $E_F$ corresponding to the Dirac point of the doped graphene substrate. The H-resonance is pinned around the Fermi level at equilibrium (0 V), but can be shifted by the tip bias.
This shows how external biases may be used to turn the spin-polarization "on/off" in the H/Gr system and how this has implications not only for the tunnel current between tip and sample (PDOS) but also affects the in-plane transport in graphene. 

\section{Model and time-dependent methods}
\begin{figure*}
    \centering
    \includegraphics[width=\textwidth]{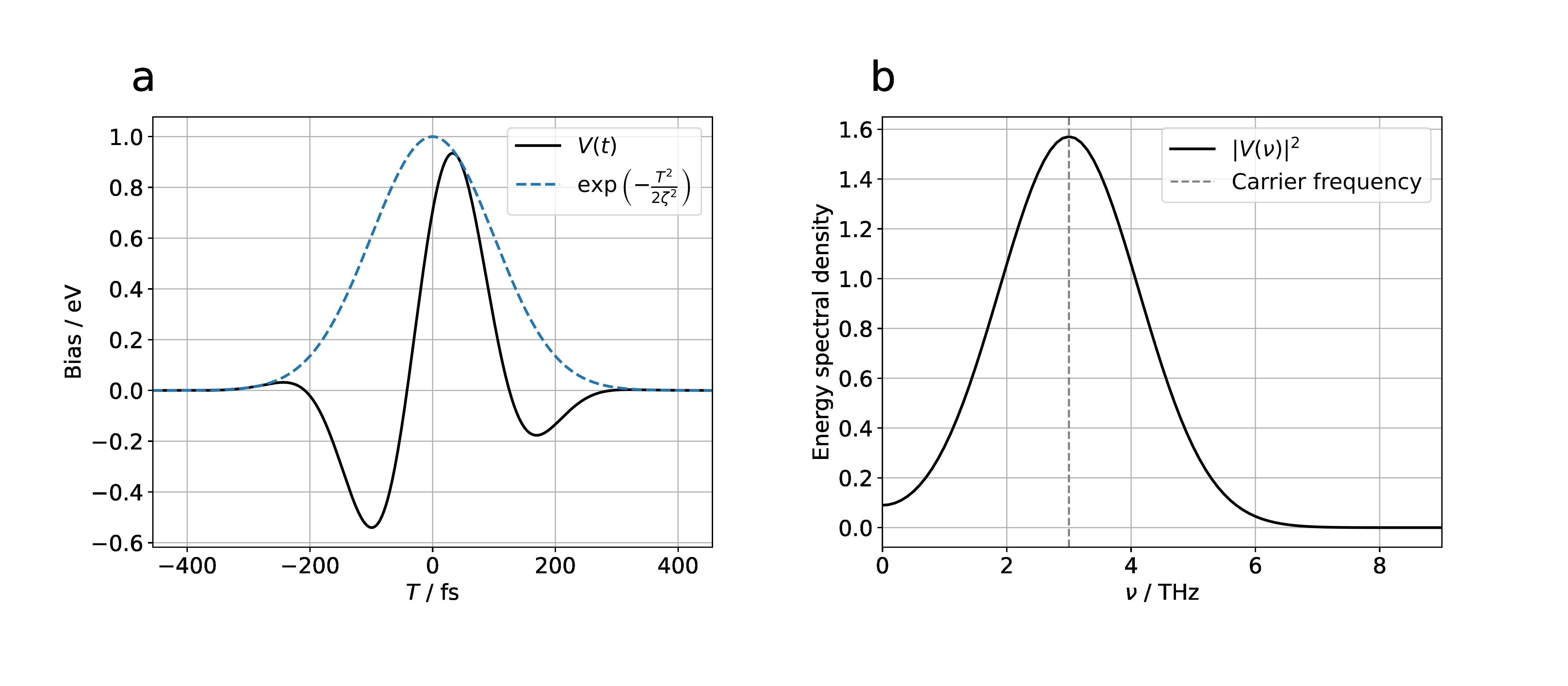}
    \caption{THz pulse used to drive the system in the time-domain (a) and frequency domain (b). The pulse has the form $V(t) = A \cos(2\pi \nu_0 t - \phi)\exp\left(-\frac{t^2}{2\zeta^2}\right)$. 
    For the numerical investigations carried out in this work, we fixed the parameters at $\zeta = \frac{3}{10} \nu_0^{-1}$, $\phi = -\frac{\pi}{4}$ and $A = 1$ eV. These parameters were chosen to mimic pulses used in experiment. The carrier frequency $\nu_0$ is varied between 1 and 10 THz; in the figure, a pulse with $\nu_0 = 3$ THz is shown.}
    \label{fig:pulse_3THz}
\end{figure*}

Consider next the response of the H/Gr system to a time-dependent variation of the tip and gate potentials with the waveform shown in Fig.~\ref{fig:pulse_3THz}a chosen to mimic THz pulses in typical experiments\cite{Ammerman2021,Peller2021}, $V(t) =  A \cos(2\pi \nu_0 t - \phi)\exp\left(-\frac{t^2}{2\zeta^2}\right)$. The time-resolved shape and power spectrum of the pulse is shown in Fig.~\ref{fig:pulse_3THz}. 
We note that a realistic experimental pulse integrates to zero over time, which is not fulfilled by the simple analytic expression specified above. However,
we do not expect this deviation to have a significant effect on the dynamics.
The central (carrier) frequency $\nu_0$ can be adjusted to investigate how the dynamics depend on the speed of external driving, thereby testing the adiabaticity in the system. The width is adjusted along with the frequency such that the dimensionless product $\zeta \nu_0 = 0.3$ remains constant.

We will use a minimal model for the H/Gr system with parameters chosen according to the DFT calculations above and model the system in the typical electrode-device-electrode setup where the time-dependence of the potential is considered in the device region, while the electrodes are assumed to be perfectly screened, so that the time dependence in these regions consists only of a rigid shift of the energies and chemical potentials following the pulse \cite{antti-book}.
The device region consists of the hydrogen atom and the sp$^3$ carbon atom in the graphene on which it is adsorbed. This system is modeled using an electronic 2-level model for each spin, corresponding to one orbital per atom per spin. The coupling to the graphene sheet and the STM tip electrodes is included using self-energies. We note that the spin-polarization in the H/Gr in reality involves carbon atoms further away from the adsorbate\cite{H}, but we will neglect these effects here.

The system is described by the total Hamiltonian,
\begin{equation}
\label{eq:general-hamiltonian}
    \hat{\mathcal{H}}(t) = \hat{H}(t) + \hat{H}_{\mathrm{gr}}(t) + \hat{H}_\mathrm{tip}(t)  + \hat{V}_{d,\mathrm{gr}} + \hat{V}_{d,\mathrm{tip}},
\end{equation}
where  $\hat{H}_{\mathrm{gr}}$ describes the graphene sheet, $\hat{H}_\mathrm{tip}$ the STM tip, $\hat{H}$ the device region, and $\hat{V}_{d,\mathrm{gr}}$, $\hat{V}_{d,\mathrm{tip}}$ the graphene-device and tip-device coupling, respectively. 
The rigid shift of the energies in the screened electrodes caused by the electric field in the THz pulse is $\Delta_\alpha V(t) $, where $\alpha \in \{\mathrm{gr}, \mathrm{tip}\}$, and we have
\begin{equation}
    \hat{H}_\alpha(t) = \sum_{k,\sigma} \left[\eps_{\alpha k} + \Delta_\alpha V(t)\right] \hat{c}\d_{\alpha k\sigma} \hat{c}\vd_{\alpha k \sigma}.
\end{equation}
 Here, $\alpha k$ is a generalized state label referring to the eigenstates of the isolated graphene sheet and tip, respectively. 
The mean-field Hubbard Hamiltonian $\hat{H}$ describing the device is,
\begin{equation}
\label{eq:H_device}
\begin{aligned}
	\hat{H}(t) = \sum_\sigma \left[\eps_H + V_\mathrm{DC} + U n\vd_{H\bar{\sigma}} (t) \right] \hat{c}\d_{H\sigma} \hat{c}\vd_{H\sigma} + \eps\vd_C(t) \hat{c}\d_{C\sigma} \hat{c}\vd_{C\sigma} 
 + (v\vd_{CH} \hat{c}\d_{C\sigma} \hat{c}\vd_{H\sigma} + \mathrm{h.c.})\,,
 \end{aligned}
\end{equation}
where $\hat{c}^{(\dagger)}_{C,H}$ are annihilation (creation) operators for the hydrogen and adsorption-site orbitals, $\eps_H$ and $\eps_C$ are the on-site energies for these orbitals, $v_{CH}$ the hopping matrix element between these, and $\sigma \in \{\up,\down\}$ refers to the spin.
The Hubbard interaction strength is given by $U$ and the opposite-spin H occupation $n\vd_{H\bar{\sigma}} = \E{\hat{c}\d_{H \bar{\sigma}} \hat{c}\vd_{H\bar{\sigma}}}$, where $\bar{\sigma}$ refers to the spin opposite to $\sigma$. The parameter $V_{\rm DC}$ designates a DC potential due the tip bias, as shown in Figs.~\ref{fig:H} and \ref{fig:CS}.
The device couples to the tip by tunneling through the hydrogen orbital, 
\begin{equation}
\hat{V}_\mathrm{d,tip} = \sum_{k\sigma} v_{\mathrm{tip},k}\,\hat{c}\d_{H\sigma} \hat{c}\vd_{\mathrm{tip},k\sigma} + \rm{h.c.},
\end{equation}
and to the graphene sheet through the C atom on the adsorption site
\begin{equation}
\hat{V}_\mathrm{d,gr} = \sum_{k\sigma} v_{\mathrm{gr},k}\, \hat{c}\d_{C\sigma} \hat{c}\vd_{\mathrm{gr},k\sigma} + \rm{h.c.},
\end{equation}
such that $v_{\mathrm{tip},k}$ is the hopping matrix element between the tip state $k$ and the hydrogen orbital, and $v_{\mathrm{gr},k}$ is the hopping matrix element between the graphene state $k$ and the device carbon orbital.
These parameters enter the numerical model only indirectly through the electrode self-energies to be specified below. 

The values of the model parameters were estimated to mimic the results of the DFT calculation, and we find that the values $\eps_H = -\SI{1.7}{eV}$, $\eps_C(t) = \SI{0}{eV} + \Delta_\mathrm{gr} V(t)$, $v_{CH} = \SI{3.25}{eV}$, $U = \SI{6.5}{eV}$, $\Delta_\mathrm{gr} = 1$, and $\Delta_\mathrm{tip} = 0$ resulted in qualitative agreement between the models when comparing PDOS on the respective atoms and their spin splitting. The effect of the THz pulse is to shift the graphene electronic energies relative to the tip.
There is no spin-flipping mechanism, such that the spin-up and spin-down electrons effectively form two separate systems, interacting only through the time-dependent mean-field occupation of the opposite spin on the H-atom, which must be solved self-consistently. 

The dynamical evolution of the system for a given time-dependence may be described using the non-equilibrium Green's function (NEGF) formalism. 
We now express all operators as matrices in the device state space, consisting of the hydrogen and carbon orbital for each spin.
The retarded, advanced, and lesser Green's functions in the device region are governed by the KBEs \cite{antti-book},
\begin{equation}
\label{eq:KBeq-retarded}
\begin{gathered}
    \left[i \p_t - H(t)\right] G^{r/a}(t,t') = \mathbf{1} \delta(t-t') 
    + \int \dd t_1 \Sigma^{r/a}(t,t_1) G^{r/a}(t_1,t'),
\end{gathered}
\end{equation}
for the retarded/advanced functions, and 
\begin{equation}
\label{eq:KBeq-lesser}
\begin{gathered}
    \left[i \p_t - H(t)\right] G^<(t,t') 
    = \int \dd t_1 \left[\Sigma^r(t,t_1) G^<(t_1,t')
    + \Sigma^<(t,t_1) G^a(t_1,t')\right]
\end{gathered}
\end{equation}
for the lesser Green's function. 
In these equations, $\Sigma = \Sigma_\mathrm{gr} + \Sigma_\mathrm{tip}$ is the self-energy related to the coupling of the device to the graphene sheet and the tip.
These self-energies may be found from the Green's function operators $g_\alpha$ of the electrodes in the absence of coupling to the device,
\begin{equation}
    \Sigma^{r/a/<}_\alpha(t,t') = \sum_k v_{\alpha,k}\vd \, g_{\alpha,k}^{r/a/<}(t,t')\, v^*_{\alpha,k}
\end{equation}
In the present work, the tip is modeled in the wide-band limit using an energy-constant, $\Sigma^{r/a}_\mathrm{tip} = \mp i \Gamma_\mathrm{tip}/2$,
with the weak tip-hydrogen coupling $\Gamma_\mathrm{tip} = 0.1$ eV,
while the self-energy of the graphene sheet is found from a tight-binding model of the graphene with nearest-neighbor hopping parameter $t = - 2.7$ eV and zero on-site energy. 
From this model the self-energy accounting for the presence of the rest of the graphene sheet on the carbon atom under the hydrogen was found using a $\mathbf{k}$-integral to obtain the real-space self-energy in the primitive unit cell of graphene \cite{papior2019removing}. Subsequently, this self-energy is further downfolded to the C atom entering the C-H system.
This yields a self-energy which reproduces the Dirac-cone DOS for the graphene.

We note that all quantities in Eqs. (\ref{eq:KBeq-retarded}-\ref{eq:KBeq-lesser}) are matrices in the device state space. Thus, in our model we obtain two sets of $2\times 2$ matrices, one set for each spin. 
To solve the KBEs, we use two different numerical approaches. The first is a method based on the separation of time-scales \cite{Honeychurch2019TimescaleFields}, where the resulting equations of motion are expanded in an asymptotic series which we denote the dynamical expansion (DE). This represents an approximation to the dynamics which becomes exact when the external time-dependence varies on a much slower time-scale than the electronic degrees of freedom in the system. 
The second method (AM) recasts the KBEs into a set of coupled ordinary differential equations (ODEs) which depend only on a single time variable. The system dynamics can then be found by propagating these ODEs forward in time from a specified initial condition \cite{Croy2009PropagationDevices,Popescu2016EfficientNanoelectronics}. This method provides an exact propagation of the dynamics, but requires that the self-energies be approximated by a series of Lorentzian functions. In the following, we will provide some more details on these two schemes. 

\subsection{Dynamical expansion}\label{subsection:DE}
We introduce a central time variable $T$,
and a difference time variable $\tau$,
\begin{equation}
T = \frac{t+t'}{2} \qc \tau = t - t'.
\end{equation}
In the absence of external driving, the system is invariant with respect to translations in time, and thus in this case the quantities in the KBEs will only be a function of the difference time $\tau$. 
The introduction of an external time-dependent driving breaks time-translation invariance and thus introduces a dependence on the central time variable $T$. Therefore, the variables $(T,\tau)$ are well-suited to separate the time-scales associated with internal dynamics of the system ($\tau$) and the external driving ($T$).
After the introduction of the new time-variables, we cast the KBEs in the so-called Wigner space representation by performing a Fourier transform over $\tau$. Thus we obtain a new set of Green's functions and self-energies
\begin{equation}
    \wt{G}(T,\omega) = \infint \dd \tau \,G(T,\tau) e^{i\omega \tau},
\end{equation}
where the superscript $\sim$ indicates that a quantity is in the Wigner representation. 
After transforming all self-energies and Green's functions to the Wigner representation, the KBEs take the form
\begin{equation}
\label{eq:KB-wigner-retarded}
\begin{gathered}
\left[\frac{i}{2} \p_T + \omega - e^{-\frac{i}{2} \p^G_\omega \p^{H}_T} H(T) \right] \wt{G}^{r/a} 
= 1 + e^{-\frac{i}{2} \left(\p^{\Sigma}_T \p^G_\omega  - \p^{\Sigma}_\omega \p^G_T \right)  }
\wt{\Sigma}^{r/a} \wt{G}^{r/a}, 
\end{gathered}
\end{equation}
and
\begin{equation}
\label{eq:KB-wigner-lesser}
\begin{gathered}
\left[\frac{i}{2} \p_T + \omega - e^{-\frac{i}{2} \p^G_\omega \p^{H}_T} H(T) \right] \wt{G}^< 
= e^{-\frac{i}{2} \left(\p^{\Sigma}_T \p^G_\omega  - \p^{\Sigma}_\omega \p^G_T  \right) }
\left(\wt{\Sigma}^< \wt{G}^a + \wt{\Sigma}^r \wt{G}^<\right),
\end{gathered}
\end{equation}
where the differential operators in the exponentials act only on the function whose superscript they bear. 
For notational convenience, we have suppressed the arguments $(T,\omega)$ on the self-energies and Green's functions.
We note that Eqs. (\ref{eq:KB-wigner-retarded}-\ref{eq:KB-wigner-lesser}) are still formally exact.
If the time-dependence is taken to be adiabatic, all central-time derivatives in Eqs. (\ref{eq:KB-wigner-retarded}-\ref{eq:KB-wigner-lesser}) vanish. 
In this limit, the exponentials act as the identity operator, and we obtain the well-known steady-state KBEs in the frequency domain
\begin{equation}
\label{eq:adiabatic-KB-wigner-retarded}
\left[ \omega - H(T) \right] \wt{G}^{r/a} _0
= 1 + 
\wt{\Sigma}^{r/a} \wt{G}^{r/a}_0.
\end{equation}
\begin{equation}
\label{eq:adiabatic-KB-wigner-lesser}
\left[ \omega - H(T) \right] \wt{G}^<_0 = \wt{\Sigma}^< \wt{G}^a_0 + \wt{\Sigma}^r \wt{G}^<_0,
\end{equation}
where the 0-subscripts indicate that these are \textit{adiabatic} Green's functions. 
If we assume that the external perturbation acts on a much slower time-scale than any internal dynamics in the system, $G$ and $\Sigma$ will be slowly varying functions of the central time $T$. 
In that case, we can expand the exponential operators in Eqs. (\ref{eq:KB-wigner-retarded}-\ref{eq:KB-wigner-lesser}) and keep only the lowest order terms on the grounds that higher order terms contain higher powers of the central time derivative $\p_T$, which must be small by virtue of the slow variation.
This allows us to expand the Green's functions in a formal series 
\begin{equation}
\label{eq:GF_expansion_in_dT}
G = G_0 + G_1 + G_2 + \cdots,
\end{equation}
where $G_i$ satisfies an equation of motion containing terms only of $i$'th order in the central time derivative $\p_T$.
Thus, $G_0$ describes the adiabatic response of the system (Eqs. \ref{eq:adiabatic-KB-wigner-retarded}-\ref{eq:adiabatic-KB-wigner-lesser}) to the time-dependent perturbation while $G_1$ describes the first-order dynamical corrections, and so on. 
The explicit expressions for $G^{r/a}$ and $G^<$ resulting from this expansion at the \nth{0} and \nth{1} orders of approximation may be seen in Supp. Mat.

Once the Green's function governing the system have been found, the time-resolved current into the device from the graphene sheet and tip, respectively, is given by \cite{antti-book},
\begin{equation}
    \label{eq:formula-for-current}
     J_\alpha(t) = \Tr C_\alpha(t),
\end{equation}
where the current matrix $C_\alpha$ is 
 \begin{equation}
 \label{eq:current-matrix}
      C_{\alpha}(t) = \int_{-\infty} ^ t \mathrm{d}t' \left[ G_{\alpha}^>(t,t')\Sigma_{\alpha}^<(t',t) -G_{\alpha}^<(t,t')\Sigma_{\alpha}^>(t',t)  \right] + \mathrm{h.c.} 
 \end{equation} 
In the Wigner space representation, the integral in Eq. (\ref{eq:current-matrix}) becomes an exponential operator, and we find the equivalent expression in terms of Wigner-space quantities,
\begin{equation}
\label{eq:wigner_formula-for-current}
\begin{gathered}
    C_\alpha(T) = \infint \frac{\dd \omega}{2\pi}
    e^{-\frac{i}{2} \left(\p_T^G \p_\omega^\Sigma - \p_\omega ^G \p_T ^\Sigma\right)} 
    \Big[ \wt{G}^>(T,\omega) \wt{\Sigma}_\alpha^<(T,\omega) - \wt{G}^<(T,\omega) \wt{\Sigma}_\alpha^>(T,\omega) \Big] + \mathrm{h.c.}
\end{gathered}
\end{equation}

We expand the current and current matrix in a series in the same manner as we did for the Green's functions,
\begin{equation}
    J_\alpha = J_{\alpha,0} + J_{\alpha,1} + J_{\alpha,2} + \cdots \qc
    C_\alpha = C_{\alpha,0} + C_{\alpha,1} + C_{\alpha,2} + \cdots,
\end{equation}
where $J_{\alpha,i} = \Tr C_{\alpha,i}$, and $C_{\alpha,i}$ satisfies an equation containing terms of order $i$ in $\p_T$. These equations, at each order, are found from the series expansion of the exponential operator in Eq. (\ref{eq:wigner_formula-for-current}) and the expansion Eq. (\ref{eq:GF_expansion_in_dT}) of the Green's functions, by collecting all terms of the same order in $\p_T$. They are given explicitly in Supp. Mat. 

It is worth emphasizing that the method does not require any propagation of the equations of motion because the dynamics are included as time-dependent corrections to the adiabatic solution. It is therefore well-suited to problems where exact propagation may not be computationally feasible, such as problems that require the consideration of dynamics over very long time scales, and, in particular, the computations may be parallelized trivially over the discrete time-steps of the central time variable, $T$. 

\subsubsection{Self-consistent solution}
The device Hamiltonian is an input parameter to the dynamical expansion, and must thus be known \textit{a priori}. This represents a problem for working with systems with interactions or electron-electron repulsion, as the effective device Hamiltonian will then depend on the state of the system. One approach to this problem is the introduction of correlation self-energies, which must then also be approximated by an additional expansion in $\p_T$ \cite{Kershaw2019Non-equilibriumInteractions}. 
In this work, we employ instead an iterative procedure, seeking a self-consistent dynamical electron density. 
This approach is in the spirit of DFT, and we envision that future implementations of the method may work with DFT programs to obtain self-consistent dynamical corrections to time-dependent problems. 
In our approach, we start from a self-consistent \textit{adiabatic} solution on a pre-defined time-grid. The self-consistent adiabatic Hamiltonian obtained from this solution is then used as the input Hamiltonian to the DE model, from which a new dynamical Hamiltonian is obtained on the same time-grid. This process is repeated until the self-consistency of the dynamical Hamiltonian is achieved. 
Obtaining convergence for all time-steps is not trivial except at very low frequencies $\nu_0$ of the external perturbation, and was not obtained at the highest frequencies investigated. 
This is related to the fact that the higher-order harmonics generated by the switching of the magnetization state are highly non-adiabatic and, therefore, unsuitable for description by an expansion such as the DE. 
Nonetheless, for frequencies below $\nu_0 \approx 5$ THz, the dynamical corrections were in good agreement with the exact time propagation (AM) scheme even when the self-consistent cycle did not formally converge for all time-steps.
Further details of the numerical implementation may be found in Supp. Mat.

\subsection{Exact Density Matrix Propagation (AM)}
The KB equation (\ref{eq:KBeq-lesser}) can be dealt with under the assumption of the spin-independent electrode broadening matrices $\Gamma_\alpha = i(\Sigma^r_\alpha - \Sigma^a_\alpha)$
being a weighted sum of Lorentzians on the form \cite{Croy2009PropagationDevices, Popescu2016EfficientNanoelectronics, xie2012time, zheng2010time}
\begin{align}
\label{eq:L_Gamma}
    \Gamma_{\alpha} (\epsilon) = \sum_{l=1}^{N_l}\frac{\gamma_{\alpha l}^2}{(\epsilon - \epsilon_{\alpha l})^2 + \gamma_{\alpha l}^2}W_{\alpha l},
\end{align} 
with a corresponding retarded self-energy
\begin{align}
\label{eq:L_SE}
    \Sigma_{\alpha}^r (\epsilon) = \frac{1}{2}\mathbb{H}[\Gamma_\alpha](\epsilon)  -\frac{i}{2}\Gamma_\alpha(\epsilon).
\end{align} 
In Eq. (\ref{eq:L_Gamma}), $W_{\alpha l}$ are fitting coefficients in a matrix sense and in Eq. (\ref{eq:L_SE}) $\mathbb{H}$ denotes the Hilbert transform. This leads to a particularly simple form of the lead self-energies  and allows the KBE to be solved exactly, in terms of the density matrix $\rho(t) = -iG^{<}(t,t)$ and a collection of auxiliary modes that contains the necessary information of the two-time Green's function. The coupled set of ODEs for propagating the density matrix then becomes \cite{Croy2009PropagationDevices,Popescu2016EfficientNanoelectronics, xie2012time,zheng2007time, jauho1994time, hu2011pade, jin2008exact}
\begin{align}
    \label{eq:AM_EOM}
    i\partial_t\rho(t) &= [H(t), \rho(t)] +i\sum_\alpha C_{\alpha}(t),
\end{align}
where the current matrix $C_\alpha$ is defined in Eq. (\ref{eq:current-matrix}). 
The memory integral is handled through contour integration and identifying so-called auxiliary modes to obtain the auxiliary mode approach to the problem \cite{Popescu2016EfficientNanoelectronics}. The current matrix $C_{\alpha}(t) $ has several associated equations of motion with coefficients related to the weights and poles in Eq. (\ref{eq:L_Gamma}) \cite{Popescu2016EfficientNanoelectronics}.
The system of ODEs \eqref{eq:AM_EOM} can then be solved using standard a ODE solver method. In this particular case, an adaptive Runge-Kutta-Fehlberg fourth order method (RKF4(5)) has been used \cite{hairer1993solving}. This serves as a basis of comparison as this method is only dependent on having a suitable fit of $\Gamma_{\alpha}$ instead of having an approximation on how fast the system can change. Because of the setup, getting the fit of $\Gamma_{\alpha}$ for each electrode is just fitting two scalar functions to the form in \eqref{eq:L_Gamma}, which is easily done using an equally spaced grid and minimizing the least squares error with respect to the fitting coefficient on each Lorentzian. 

\begin{figure}
    \centering
    \includegraphics[width=\textwidth]{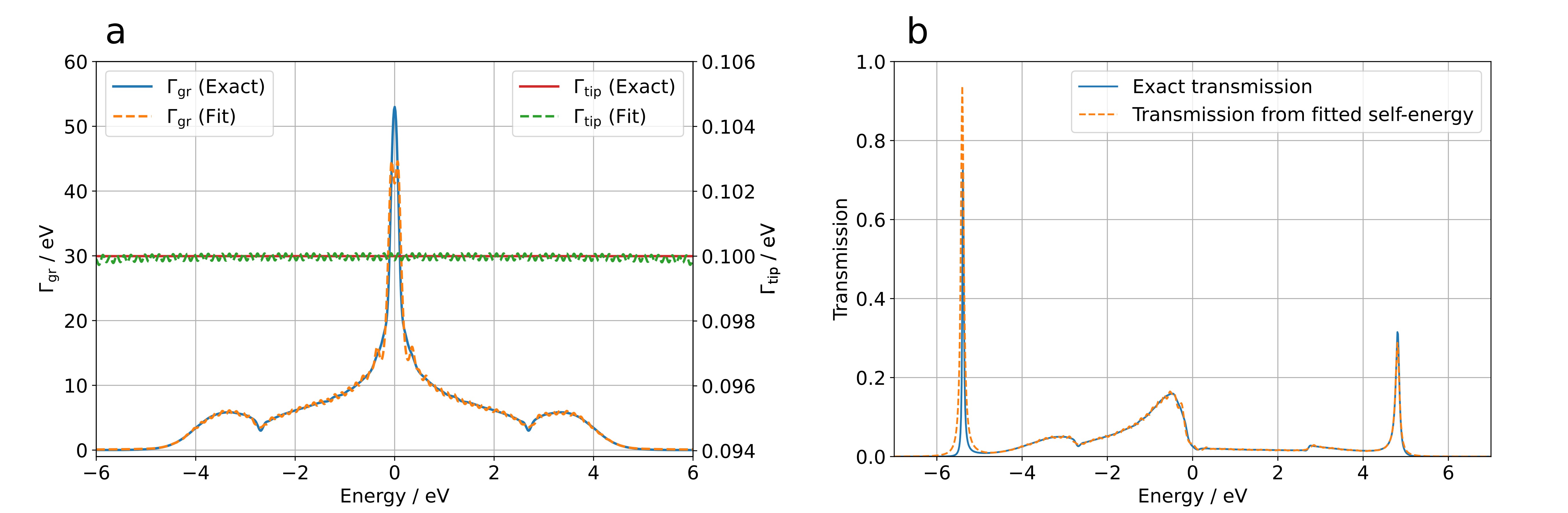}
    \caption{a) Lorentzian fits to $\Gamma_\mathrm{gr}$ (blue) and $\Gamma_\mathrm{tip}$ (red) used for the AM approach. b) Tip-graphene transmission function calculated with the numerically "exact" self-energies and the fitted $\Gamma$'s without the Hubbard term ($U = 0$). The transmission was calculated using TBtrans \cite{Papior2017ImprovementsTransiesta}. }
    \label{fig:fits}
\end{figure}

These steps, from obtaining fits to the lead $\Gamma_\alpha$'s to propagating the density matrix and its auxiliary modes in time in a numerically exact way, have been implemented in a code that will be published in the future \cite{FutureTDPaper}. The fits used can be seen in Fig.~\ref{fig:fits}a together with a comparison of the transmission function (tip to graphene) with numerically "exact" self-energies and fitted self-energies in Fig. \ref{fig:fits}b. For the fits, as shown in figure \ref{fig:fits}, 100 Lorentzians were used.

\begin{figure}[htpb]
    \centering
    \includegraphics[width=\linewidth]{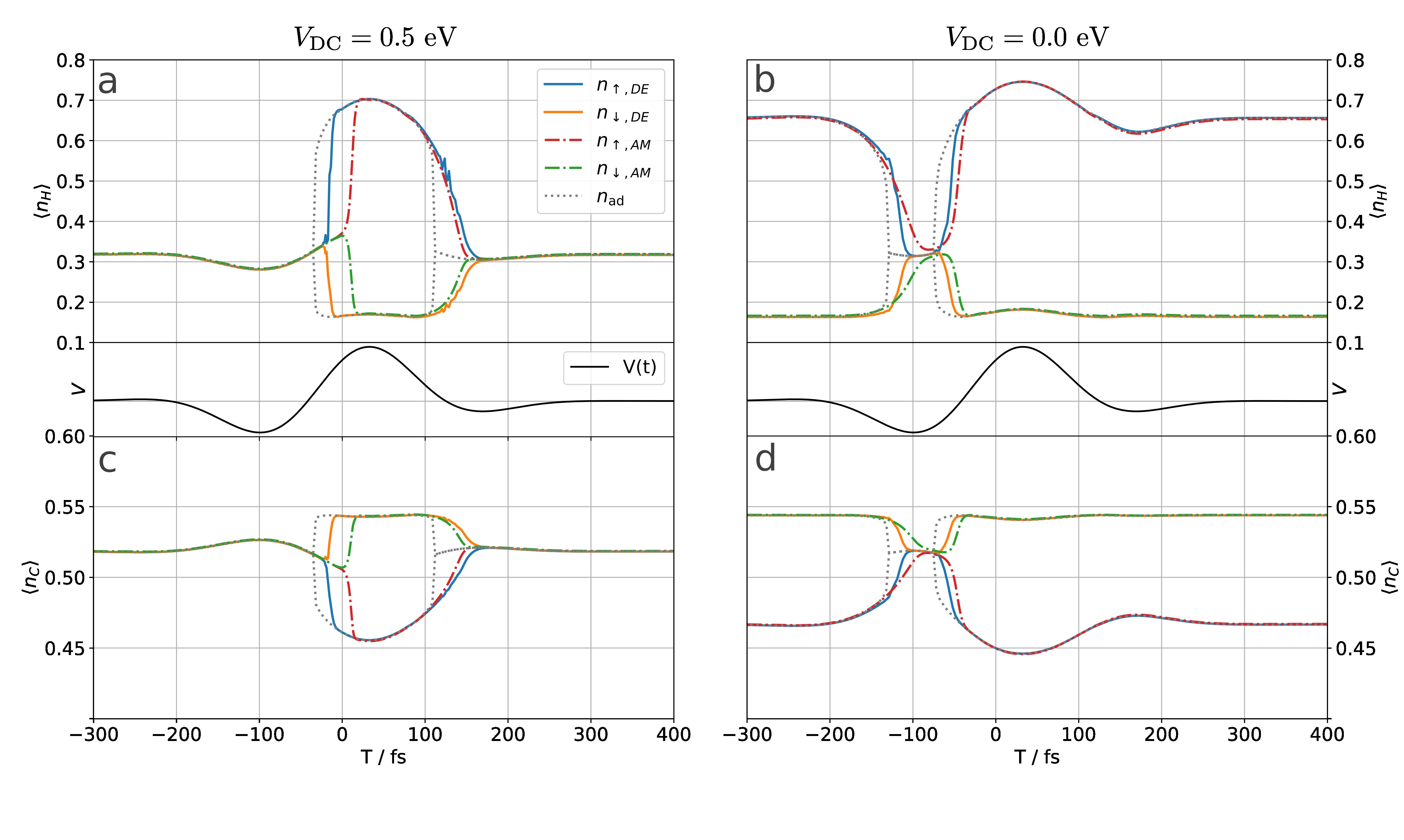}
    \caption{a) Time-dependent occupation of the hydrogen orbital ($\langle n_H \rangle(t)$) as determined self-consistently by the dynamical expansion (DE) and the auxiliary mode (AM) method when the system is perturbed by a THz pulse at a central frequency of 3 THz (middle panels) for a DC bias where the system is non-polarized ($V_{\rm DC}=0.5\,\rm{eV}$).
    b) H-occupation for a DC bias yielding a polarized system ($V_{\rm DC}=0\,\rm{eV}$). c) Carbon $p_z$-orbital occupation ($\langle n_C \rangle(t)$) for non-polarized, and d) polarized cases.
    In each case, the lines labeled $n_\mathrm{ad}$ marks the adiabatic solution to the problem.} 
    \label{fig:densities-3THz}
\end{figure}

\section{Results}
To test our methods, we calculate self-consistently the time-dependent occupation of spin-up and spin-down electrons on the H atom and central C atom, along with the time-resolved spin-polarized current injected into the graphene sheet. A net current may be injected since electrons can tunnel from the STM tip through the hydrogen atom and into the graphene sheet. These properties are calculated using the two methods outlined above. For our purposes, the AM method may be regarded as numerically exact. In the DE method, the equations of motion are expanded to first order, i.e. we include only first-order dynamical corrections. In the fully adiabatic limit the self-consistent solution to the electronic dynamics may be found semi-analytically, and for comparison we include such a calculation in all figures shown.

In the absence of the THz pulse, the system relaxes to an equilibrium configuration which is similar to the ground-state found in the DFT calculations in Section 3. In particular, the ground state is spin-polarized when $V_\mathrm{DC}= 0$, but the polarization vanishes upon the inclusion of a sufficiently large bias. We analyze in the following the dynamics 
 for $V_\mathrm{DC}=0.5$ eV and $V_\mathrm{DC}= 0.0$ eV. These two situations correspond to the situations in the top (0.4 V) and bottom (0.0 V) panels in the left $(g=0)$ part of Fig. \ref{fig:H}c, respectively. 
 This way of applying a DC bias on the system mimics that used in the experiment by Cocker et al. \cite{Cocker2016}.

In Fig.~\ref{fig:densities-3THz} we show the time-dependent occupation of the hydrogen (a, b) and carbon (c, d) orbitals as determined self-consistently by the dynamical expansion (DE) and the auxiliary mode (AM) method when the system is perturbed by the THz pulse. We take the central frequency of the pulse to be $\nu_0 = 3$ THz and its amplitude to 1 eV as in the experiment by Peller and co-workers \cite{Peller2021}. We will show in the following that the DE method breaks down for this system between 5 and 10 THz, and so 3 THz is one of the highest frequencies for which we would expect the method to work.
The gray dotted line marks the adiabatic self-consistent solution, which neglects all dynamics associated with the finite response time of the electronic system.
The main new feature when dynamics are included is a delay in the switching time from the polarized to non-polarized state, and vice versa.
The DE and AM methods are in reasonably good agreement with respect to this time delay, in particular for the polarized $\ra$ non-polarized (P $\ra$ NP) transition for the DC biased system in Fig. \ref{fig:densities-3THz}a and c. 
Note that for the NP $\ra$ P transition to occur, the spin-up/down symmetry of the system must be spontaneously broken. 
Physically this is achieved by any random (e.g. thermal) fluctuation of the system, but in the numerical implementation, a small spin-up/down asymmetry is introduced explicitly to break this symmetry. We implemented this by introducing a small energy splitting between spin up and down electrons, $\eps_H \ra \eps_H + \eps_\sigma$, where $\eps_{\up/\down} = \mp \SI{1}{\micro\electronvolt}$, i.e. a much smaller energy scale than any other scale in the system. 
This parameter therefore controls the spontaneous symmetry breaking during the NP $\ra$ P transition without otherwise affecting the dynamics. 

The sign of $\eps_\sigma$ has been chosen to make the system eventually return to the initial polarization state for the unbiased configuration. One could equally well have chosen the opposite sign, which would make the pulse switch the polarization state of the H adatom for sufficiently low central frequencies.
In the absence of an external magnetic field splitting the degeneracy between the spin states, and if the central frequency of the pulse is sufficiently low to completely quench the polarization, the final polarization state would be independent of the initial state.

In Fig.~\ref{fig:densities-3THz}c we see that as the hydrogen atom acquires a spin polarization, the carbon atom acquires the reverse polarization.
This happens even though the model does not include e-e repulsion in the carbon orbital. 
It is known that the spin-polarization of a hydrogen adsorbate on graphene induces a sublattice-dependent spin polarization in the graphene, extending for several unit cells around the adsorption site \cite{H,Gao2021ControlGating}. 
Our observation is qualitatively in line with these results, though the effect could be modeled more accurately by using a spin-dependent coupling self-energy for the graphene-device contact. One could also include more carbon atoms in the device region and thus explicitly model their time-dependent spin-polarization. 
The AM results for the carbon orbital have been shifted by a constant vertical off-set to make all methods agree in the steady-state regime. If such an offset is not included, a small discrepancy is observed due to the Lorentzian fit to the self-energy used in the AM method in place of the exact self-energy. The carbon orbital, which couples directly to the graphene sheet through the self-energy, is particularly sensitive to any variation in this parameter. We remark that the quality of the fit may always be improved by including more Lorentzian functions, until a satisfactory fit is achieved. 

\begin{figure*}[htpb]
    \centering
    \includegraphics[width=\linewidth]{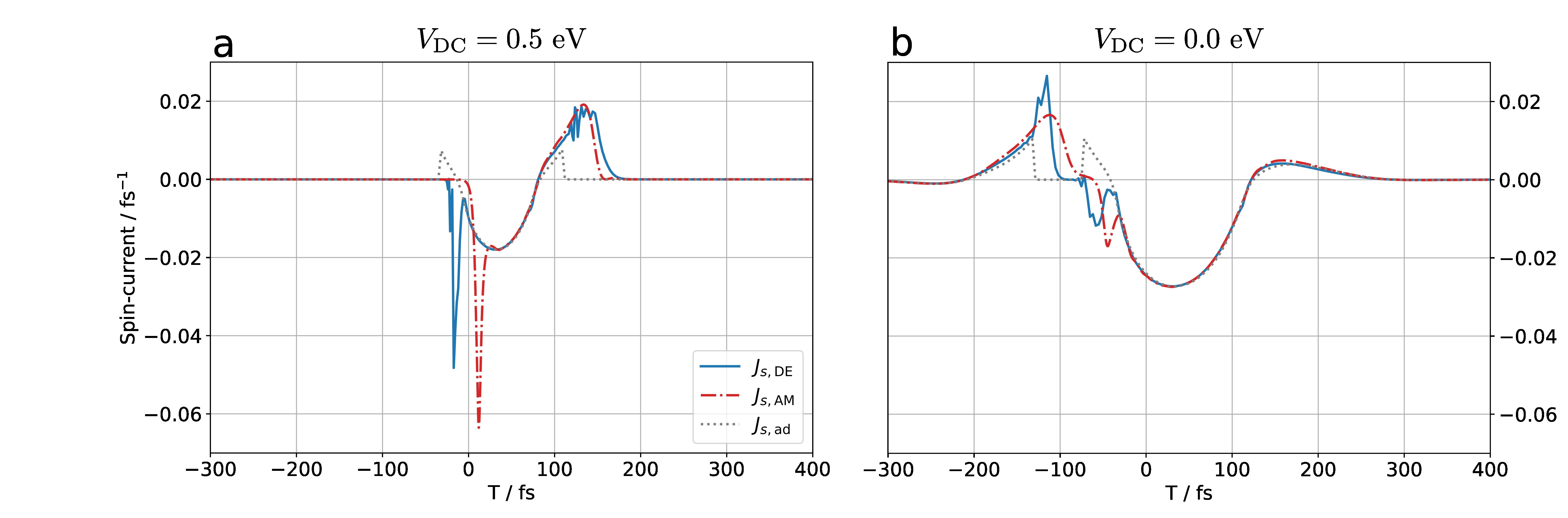}
    \caption{Spin-polarized current $J_s= J_\up - J_\down$ into the graphene layer for (DC) biased (a) and unbiased (b) configurations. A positive sign of $J_s$ means that there is a net spin-up current into the graphene sheet. Both time-dependent methods predict large spikes in the current when the polarization state of the hydrogen atom changes, in particular during the transition from non-polarized to polarized. This feature is absent in the adiabatic calculation.}
    \label{fig:spin_current_3THz}
\end{figure*}

Figure \ref{fig:spin_current_3THz} shows the time-resolved spin-polarized current $J_s = J_\uparrow - J_\downarrow$ injected into the graphene due to the THz pulse for the DC biased and unbiased configurations.
Large peaks are observed when the magnetization state switches, corresponding to the abrupt change in occupation of the hydrogen atom. 
These peaks are completely absent in the adiabatic calculation. The DE result displays rapid small-amplitude oscillations shortly before the switch occurs. This is a numerical artifact and reflects the fact that the self-consistent cycle did not converge for these particular time-steps. 
This is not surprising, as the DE approach is based on the assumption that the time-evolution be near adiabatic, which is far from the case near the transition, see the discussion in Supp. Mat.
Despite this challenge, the overall prediction by the DE method is in good agreement with the exact (AM) result, especially for the biased (NP $\ra$ P $\ra$ NP) configuration.

\begin{figure*}[htpb]
        \centering
    \includegraphics[width=\linewidth]{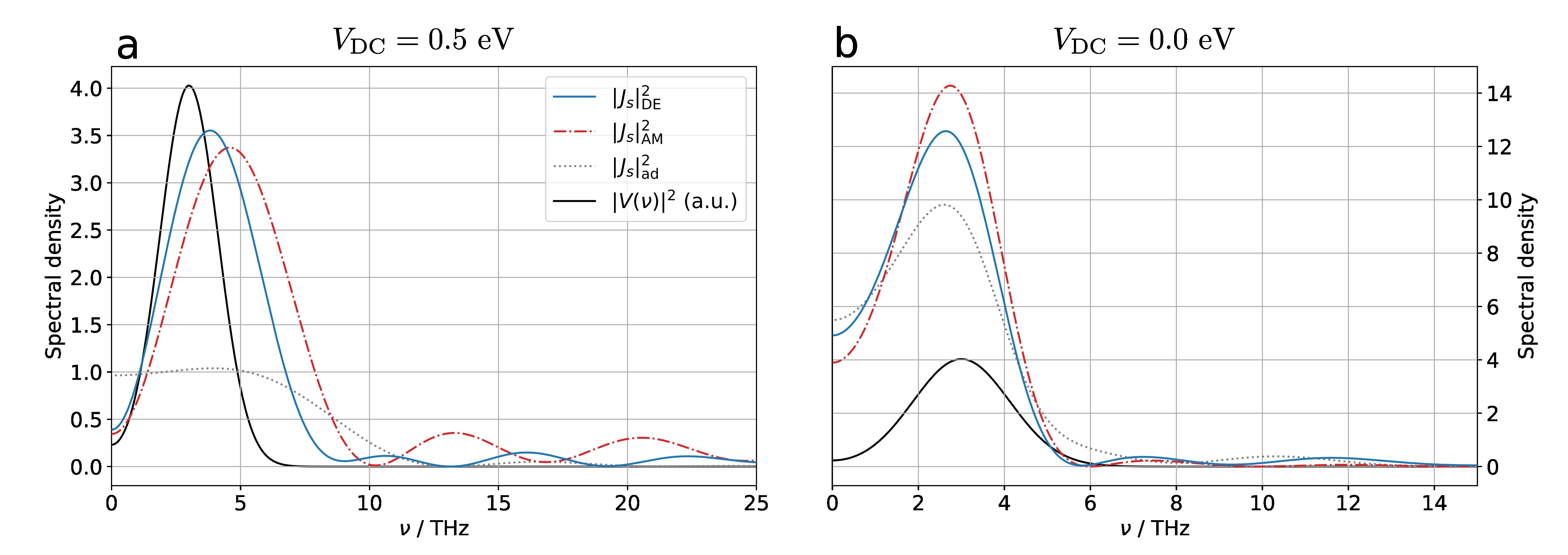}
    \caption{Spectrum of the spin-polarized current $J_s= J_\up - J_\down$ into the graphene layer for (DC) biased (a) and unbiased (b) configurations. Due to the spike in current associated with the spin-polarization phase transition, the spectrum of the current contains much larger frequencies than that of the external pulse, an indication that the problem is highly non-linear. This is especially evident in the biased case, when the non-polarized$\rightarrow$ polarized transition is particularly rapid. Note the different scales of both axes in the two figures; the black curve showing the spectrum of the external pulse in arbitrary units is the same in both figures. }
    \label{fig:spectrum_3THz}
\end{figure*}

The spectrum of the spin-polarized current may be seen in Fig. \ref{fig:spectrum_3THz}. 
The DE and AM spin-polarized currents contain significant Fourier components at much higher frequencies than the adiabatic current, as one may have anticipated from the peaks observed in Fig.~\ref{fig:spin_current_3THz}.
They both also predict oscillations in the spectra at high frequencies (most evident in Fig. \ref{fig:spectrum_3THz}a), although the amplitudes and exact locations of these oscillations are different. 
This finding indicates that the DE method captures well the qualitative effects of the dynamics, but may not be quantitatively correct on all accounts.
All three methods predict currents whose spectra contain higher frequencies than the incident THz pulse, attesting to the highly non-linear behavior of the system.

\begin{figure}[htpb]
  \centering
 \includegraphics[width=\linewidth]{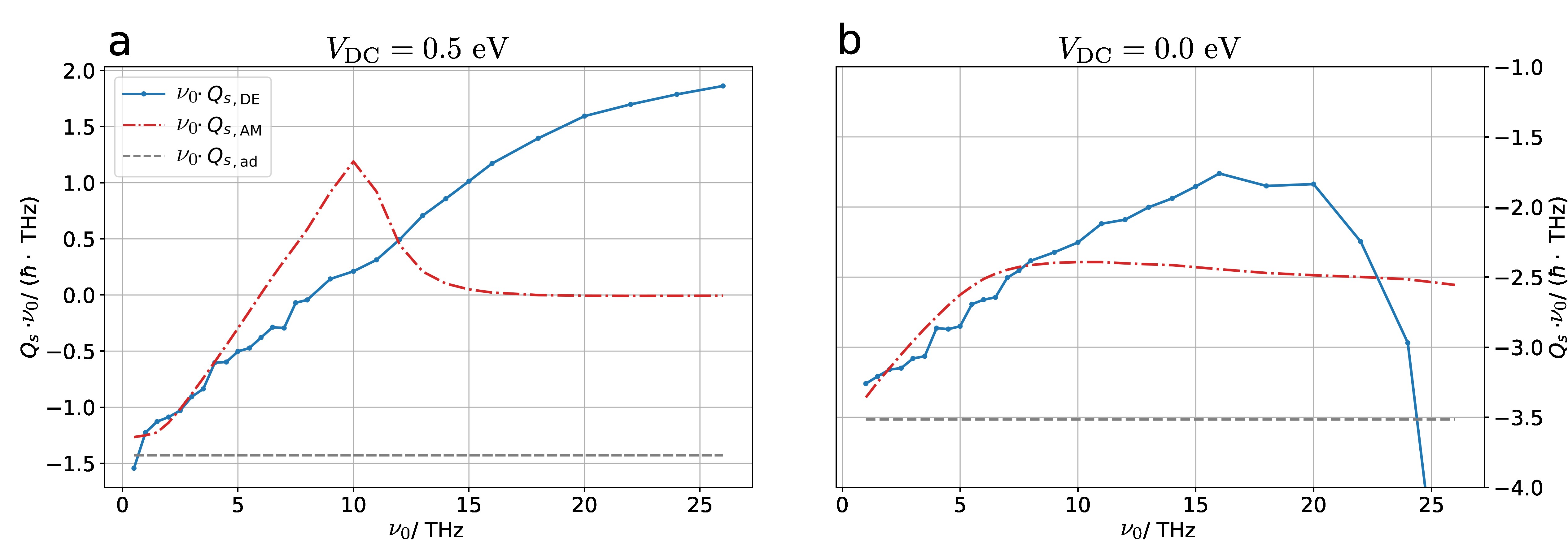}
  \caption{Total spin injected into the graphene layer multiplied by pulse frequency plotted for various frequencies for (DC) biased (a) and unbiased (b)
   conﬁgurations. The total spin injected is calculated as $Q_s = \frac12 \int (J_s) \dd t$ with $J_s$ given by the three different methods. Note the different scales of the y-axes on the two figures. 
   In the adiabatic case, the transferred charge is simply inversely proportional to the frequency, and so $\nu_0 Q_{s, \mathrm{ad}} = $ const. Multiplying by the frequency thus emphasizes the dynamical variation of the transferred charge as the frequency of the external potential is varied. }
    \label{fig:DeltaQ}
\end{figure}

As the DE method is exact in the adiabatic limit, a natural question is how fast the externally imposed time-dependence may be before the method breaks down. 
Fig. \ref{fig:DeltaQ} shows the total spin injected into the graphene sheet $Q_s(\nu_0) = \frac12 \infint J_s(t,\nu_0) \dd t$ due to a single THz pulse as a function of the central frequency $\nu_0$ of the pulse. 

Neglecting dynamical effects (i.e. the adiabatic case), the transferred spin is inversely proportional to the central frequency of the pulse because the current is frequency-independent. To highlight the difference due to the dynamical corrections, we plot $\nu_0 Q_s$ which is constant for the adiabatic calculation, and shown in Fig. \ref{fig:DeltaQ} as a gray dashed line. 
At low frequencies, the DE and AM methods are in good agreement, predicting that the transferred charge is roughly independent of $\nu_0$, which is seen in figure \ref{fig:DeltaQ} as a linear relationship between $\nu_0 Q_s$ and $\nu_0$. At low frequency, the system has time to completely polarize / depolarize , resulting in an approximately frequency-independent current associated with the filling and emptying of the device orbitals. 
As the frequency increases the DE method begins to deviate.
From the figures, we assess that $\nu_0 \approx 5$ THz is the largest frequency for which the DE works well. 
We remark that at high frequencies, the self-consistent cycle used in the DE method does not converge well. For these frequencies, the dynamical density with the smallest error obtained in the self-consistent cycle is included in the figure. The dots show the frequencies at which calculations were made; the lines interpolate between these points. 
In the biased case (Fig. \ref{fig:DeltaQ}a), the dynamical behavior predicted by the AM method changes significantly around $\nu_0 = 10$ THz. 
This change is related to the fact that at high frequencies, the electronic system cannot adjust to follow the pulse. Thus, when the system starts out in the NP state, it does not have time to polarize. For a fully non-polarized system, the spin-polarized charge will always vanish due to the spin-up/down symmetry. This explains the sudden decrease in transferred charge at this point.  This highly non-adiabatic effect is not at all captured by the DE method.  

\section{Conclusion and Outlook}
We have tested the computationally efficient dynamical expansion (DE)/time-scale separation method, introduced by \cite{Honeychurch2019TimescaleFields} on a model of an experimentally relevant system, namely mean-field, open-system calculations of charge and spin dynamics in a THz-STM junction
involving a hydrogen adsorbate on graphene.
We demonstrated that for low THz frequencies, the DE method gives a good description of the non-adiabatic dynamics compared to full propagation using the auxiliary mode (AM) method.  This calculation may furthermore be seen as a toy-model for a self-consistent TD-DFT-NEGF setup, where dynamical corrections to the density due to an external field are taken into account via the dynamical expansion method. Some of the numerical challenges and methods are discussed in Supp. Mat. 

We have shown how the THz pulse generates spin-dynamics in the hydrogen on graphene THz-STM system with higher harmonic frequencies appearing due to the highly nonlinear system. This dynamics may affect both the tunnel current from the tip electrode, as well as the spin-dependent scattering of carriers in graphene, as seen by the spin-dependent scattering cross section of the hydrogen adsorbate.

The magnetic structure in the graphene induced by the polarized hydrogen atom is long-ranged \cite{H}.
As an outlook it would be interesting to extend the computational region, and possibly to include dynamics among multiple localized spins. 
We also note that the carrier-envelope phase (CEP)  may be changed so the THz wave-form changes in a continuous way e.g. from maximum being positive to negative \cite{Ludwig2020}, as a further way of tuning the dynamics. 

The expansion in powers of the central time derivative $\p_T$ used in the DE may in principle be continued up to arbitrary order to include dynamical effects to higher degrees of accuracy. It would be interesting to extend the numerical implementation beyond the first order dynamical corrections. 
In a preliminary study, we found that it was difficult to converge the self-consistent cycle within the second-order DE, and we did not pursue this further.
An open question for the DE method is the issue of convergence. 
It is not trivial to quantify how slow the external time-variation in a given system should be in order for the lowest order terms to approximate the full dynamics. A further investigation into this issue would be highly interesting, in particular in combination with the inclusion of higher order terms in the formal series expansion.

\section*{Author contributions}
JKS implemented the DE method and performed the DE and
some of the AM calculations. ABL implemented the AM method
and performed some of the AM calculations. FG performed the DFT-NEGF calculations. All authors contributed to the article and
approved the submitted version.

\section*{Acknowledgements}
The authors thank Profs. Peter Uhd Jepsen and Peter B{\o}ggild for stimulating discussions. JKS, FG were supported by Villum fonden (Project No. VIL 13340), ABL by the Independent Research Fund Denmark (Project No. 0135-00372A), and APJ by the Center for Nanostructured Graphene (CNG), which is sponsored by the Danish National Research Foundation (Project No. DNRF103).

\printbibliography

\end{document}